\documentclass[12pt]{article}%
\usepackage{amsmath,latexsym}
\usepackage{graphicx}
\usepackage{amsmath}
\usepackage{amsfonts}
\usepackage{amssymb}%
\begin{document}

\title{{A viable wormhole model in a five-dimensional 
       spacetime}}
   \author{
Peter K. F. Kuhfittig*\\  \footnote{kuhfitti@msoe.edu}
 \small Department of Mathematics, Milwaukee School of
Engineering,\\
\small Milwaukee, Wisconsin 53202-3109, USA}

\date{}
 \maketitle

\begin{abstract} \noindent This paper generalizes two 
of the author's earlier wormhole solutions that are 
characterized by an extra spatial dimension.  This paper 
adds the assumption that the components of the line element  
are functions not only of the radial coordinate $r$ but of 
the extra coordinate $l$ as well, resulting in a 
significant generalization.  It is shown that the throat 
of the wormhole can be threaded with ordinary matter and 
that the unavoidable violation of the null energy condition 
can be attributed to the extra dimension.  To be consistent 
with string theory, we also assume that this extra dimension 
has a small magnitude. 

\end{abstract}


\section{Introduction}\label{S:Introduction}
While wormholes are as good a prediction of Einstein's theory
as black holes, they are subject to severe restrictions from 
quantum field theory \cite{MT88, LO09, pK23, pK20}.  In 
particular, to hold a wormhole open requires the need to 
violate the null energy condition, calling for the existence 
of ``exotic matter."  It has been shown, however, that this 
requirement can be met by assuming the existence of an 
extra spatial dimension.  Some of these issues were discussed 
previously by the author \cite{pK18, pK24}: Ref. \cite{pK18} 
assumes that the line element has an extra term of the form 
$e^{2\mu(r,l)}dl^2$, where $r$ is the usual radial coordinate, 
while $l$ is the extra coordinate.  Ref. \cite{pK24} assumes 
an extra time-dependent term of the form $e^{2\mu(r,l,t)}dl^2$. 
In this paper, the extra term has the simpler form 
$[\mu(r,l)]^2dl^2$; in addition, the redshift and shape 
functions are also functions of $r$ and $l$.  It is shown 
once again that the throat of the wormhole could be lined 
with ordinary matter, while the extra dimension is then 
responsible for the unavoidable energy violation.  

To that end, let us first recall the static and spherically 
symmetric form of a wormhole in Schwarzschild coordinates 
\cite{MT88}:  
\begin{equation}\label{E:line1}
  ds^2=-e^{2\Phi(r)}dt^2 +e^{2\lambda(r)}dr^2+r^2
  (d\theta^2+\text{sin}^2\theta\,d\phi^2),
\end{equation}
where
\begin{equation}
   e^{2\lambda(r)}=1-\frac{b(r)}{r}.
\end{equation}
Here $\Phi=\Phi(r)$ is called the \emph{redshift function} 
and $b=b(r)$ is called the \emph{shape function} since it 
determines the spatial shape of the wormhole when viewed, 
for example, in an embedding diagram.  A strong advocate 
of retaining an extra spatial dimension had been Paul 
Wesson \cite{WP92}.  The main reason is that the field 
equations for a five-dimensional totally flat spacetime 
yield the Einstein field equations in four dimensions 
containing matter, also called the \emph{induced-matter 
theory}.  It can be argued that our understanding of 
four-dimensional gravity is greatly enhanced by assuming 
a fifth dimension.  We therefore need to construct a 
model that is consistent with our knowledge of general 
relativity.  To that end, we are going to let $l$ denote 
the extra coordinate and assume that there is an 
additional term having the form $[\mu(r,l)]^2dl^2$, 
thereby retaining the dependence on the $r$-coordinate.
To make the model as general as possible, we also assume 
that the first two terms in line element (\ref{E:line1})  
are functions of $r$ and $l$, as well.  So the line 
element becomes 
\begin{equation}\label{E:line3}
  ds^2=-e^{2\Phi(r,l)}dt^2 +e^{2\lambda(r,l)}dr^2+r^2
  (d\theta^2+\text{sin}^2\theta\,d\phi^2)+[\mu(r,l)]^2 dl^2,
\end{equation}
where $e^{2\lambda(r,l)}=1-b(r,l)$.  This line element is 
more general than those discussed in Refs. \cite{pK18, pK24}. 
To be consistent with string theory, we also assume that the 
coefficient $\mu(r,l)$ has a small magnitude.

\section{Basic calculations}
To study the effects of our extra spatial dimension, we 
start with line element (\ref{E:line3}) and choose an 
orthonormal basis $\{e_{\hat{\alpha}}\}$ which is dual to
the following 1-form basis:
\begin{equation}\label{E:oneform1}
    \theta^0=e^{\Phi(r,l)}\, dt,\quad \theta^1=
    \left[1-\frac{b(r,l)}{r}\right]^{-1/2}\,dr,\\
     \quad\theta^2=r\,d\theta, \quad
        \theta^3=r\,
   \,\text{sin}\,\theta\,d\phi,\quad \theta^4=\mu(r,l)dl.
\end{equation}
These forms yield
\begin{equation}\label{E:oneform3}
     dt=e^{-\Phi(r,l)}\,\theta^0,\quad dr=
     \left[1-\frac{b(r,l)}{r}\right]^{1/2}\,\theta^1,\\
     \quad d\theta=\frac{1}{r}\theta^2, \quad
    d\phi=\frac{1}{r\,\text{sin}\,\theta}\theta^3, \quad
    dl=\frac{1}{\mu(r,l)}\,\theta^4.
\end{equation}  

To obtain the components of the Riemann curvature tensor, 
we need to determine the curvature 2-forms.  Here we use 
the method of differential forms, following Ref. 
\cite{HT90}.  To that end, we calculate the exterior 
derivatives in terms of $\theta^i$, starting with
\begin{equation}\label{E:zero}
   d\theta^0=\frac{\partial\Phi(r,l)}{\partial r}
   \left(1-\frac{b(r,l)}{ r}\right)^{1/2}\theta^1\wedge\theta^0
   +\frac{\partial\Phi(r,l)}{\partial l}\frac{1}{\mu(r,l)}
   \theta^4\wedge\theta^0.
\end{equation}
As noted in the Introduction, to be consistent with string 
theory, we need to assume that $\mu(r,l)$ is extremely 
small. So $d\theta^0$ in Eq. (\ref{E:zero}) becomes 
physically unacceptable unless
\begin{equation}\label{E:phi} 
   \frac{\partial\Phi(r,l)}{\partial l}\equiv 0.
\end{equation} 
A related problem in Eq. (\ref{E:zero}) is the shape 
function.  Momentarily returning to Eq. (\ref{E:line1}), 
let us recall that in a Morris-Thorne wormhole, the 
spherical surface $r=r_0$ is called the \emph{throat} 
of the wormhole, where $b(r_0)=r_0$ and $b(r)<r$ for 
$r>r_0$, while $b'(r_0)<1$.  In line element (\ref{E:line3}),
the first of these conditions becomes $b(r_0,l)=r_0$ for 
every $l$, which does not appear to be a natural 
requirement.  We are thereby forced to assume that the 
shape function is a function of $r$ alone.  The form of 
$d\theta^0$ therefore becomes [in view of Eq. 
(\ref{E:phi})] 
\begin{equation}   
    d\theta^0=\frac{\partial\Phi(r,l)}{\partial r}
   \left(1-\frac{b(r)}{ r}\right)^{1/2}\theta^1\wedge\theta^0.
\end{equation}

The remaining exterior derivatives are listed next:
\begin{equation} 
   d\theta^1=0,
\end{equation}
\begin{equation} 
    d\theta^2=\frac{1}{r}\left[1-\frac{b(r)}{r}\right]^{1/2}
    \theta^1\wedge\theta^2,
\end{equation}
\begin{equation}
     d\theta^3=\frac{1}{r}\left[1-\frac{b(r)}{r}\right]^{1/2}
     \theta^1\wedge\theta^3+\frac{1}{r}\text{cot}\,\theta
     \,\,\theta^2\wedge\theta^3,
\end{equation}
\begin{equation}
   d\theta^4=\frac{\partial \mu(r,l)}{\partial r}
   \frac{1}{\mu(r,l)}
   \left[1-\frac{b(r)}{r}\right]^{1/2}
  \,\theta^1\wedge\theta^4.
\end{equation}
The connection 1-forms $\omega^i_{\phantom{i}\,\,k}$ have the
symmetry
    $\omega^0_{\phantom{i}\,\,i}=\omega^i_{\phantom{0}0}
    \;(i=1,2,3,4)$\;\text{and}\;$\omega^i_{\phantom{j}j}=
     -\omega^j_{\phantom{i}\,i}\;(i,j=1,2,3,4, i\ne j)$
and are related to the basis $\{\theta^i\}$ by
\begin{equation}
   d\theta^i=-\omega^i_{\phantom{k}k}\wedge\theta^k.
\end{equation}
The solution of this system is
\begin{equation}
   \omega^0_{\phantom{0}1}=\frac{d\Phi(r,l)}{dr}
   \left[1-\frac{b(r)}{r}\right]^{1/2}\theta^0,
\end{equation}
\begin{equation}\label{E:omega}
   \omega^4_{\phantom{0}1}=
   \frac{\partial\mu(r,l)}{\partial r}
      \left[1-\frac{b(r)}{r}\right]^{1/2}
      \frac{1}{\mu(r,l)}\theta^4,  
\end{equation}
\begin{equation}
   \omega^3_{\phantom{0}2}=
   \frac{1}{r}\,\text{cot}\,\theta\,\,\theta^3,
\end{equation}
\begin{equation}
    \omega^3_{\phantom{0}1}=
    \frac{1}{r}\left[1-\frac{b(r)}{r}\right]^{1/2}\theta^3,
\end{equation}
\begin{equation}
   \omega^2_{\phantom{0}1}=
   \frac{1}{r}\left[1-\frac{b(r)}{r}\right]^{1/2}\theta^2,
\end{equation}
\begin{equation}
 \omega^0_{\phantom{0}4}=\frac{\partial\Phi(r,l)}{\partial l}
   \frac{1}{\mu(r,l)}\theta^0=0;
\end{equation}
($\omega^0_{\phantom{0}4}=0$ thanks to Eq. ({\ref{E:phi}).)
Finally,
\begin{equation}
   \omega^0_{\phantom{0}2}=\omega^0_{\phantom{0}3}=
   \omega^2_{\phantom{0}4}=\omega^3_{\phantom{0}4}=0.
\end{equation}   

The curvature 2-forms $\Omega^i_{\phantom{j}j}$ are calculated
directly from the Cartan structural equations
\begin{equation}
    \Omega^i_{\phantom{j}j}=d\omega^i_{\phantom{j}j} +\omega^i
     _{\phantom{j}k}\wedge\omega^k_{\phantom{j}j}.
\end{equation}
The results for $\Omega^i_{\phantom{j}j}$ are given in Appendix A.

The components of the Riemann curvature tensor can be read off
directly from the form
\begin{equation}
   \Omega^i_{\phantom{j}j}=-\frac{1}{2}R_{mnj}^{\phantom{mnj}i}
    \;\theta^m\wedge\theta^n
\end{equation}
and are listed next:
\begin{multline}
   R_{011}^{\phantom{000}0}=-\frac{1}{2}\frac{\partial\Phi(r,l)}
   {\partial r}\frac{rb'(r)-b(r)}{r^2}+\frac{\partial^2\Phi(r,l)}
   {\partial r^2}\left(1-\frac{b(r)}{r}\right)\\
   +\left(\frac{\partial\Phi(r,l)}{\partial r}\right)^2
   \left(1-\frac{b(r)}{r}\right),
\end{multline}
\begin{equation}
   R_{022}^{\phantom{000}0}=R_{033}^{\phantom{000}0}=
   \frac{1}{r}\frac{\partial\Phi(r,l)}{\partial r}
   \left(1-\frac{b(r)}{r}\right),
\end{equation}
\begin{equation}
   R_{044}^{\phantom{000}0}=\frac{\partial\Phi(r,l)}{\partial r}
   \frac{\partial\mu(r,l)}{\partial r}\left(1-\frac{b(r)}{r}\right )
   \frac{1}{\mu(r,l)},
\end{equation}
\begin{equation}
   R_{122}^{\phantom{000}1}=R_{133}^{\phantom{000}1}=
   -\frac{1}{2}\frac{rb'(r)-b(r)}{r^3},
\end{equation}
\begin{equation}
   R_{144}^{\phantom{000}1}=\left [\frac{\partial^2\mu(r,l)}{\partial r^2}
   \left(1-\frac{b(r)}{r}\right)-\frac{1}{2}\frac{\partial\mu(r,l)}
   {\partial r}\frac{rb'(r)-b(r)}{r^2}\right]\frac{1}{\mu(r,l)},
\end{equation}
\begin{equation}
   R_{233}^{\phantom{000}2}=-\frac{b(r)}{r^3},
\end{equation}
\begin{equation}
   R_{244}^{\phantom{000}2}=R_{344}^{\phantom{000}3}=
   \frac{1}{r}\frac{\partial\mu(r,l)}{\partial r}
      \left(1-\frac{b(r)}{r}\right)\frac{1}{\mu(r,l)}.
\end{equation}

The last form to be derived in this section is the Ricci
tensor, which is obtained by a trace on the Riemann
curvature tensor:
\begin{equation}\label{E:trace}
   R_{ab}=R_{acb}^{\phantom{000}c}.
\end{equation}
The components are listed in Appendix B.  These 
forms will come into play in the next section.

\section{The main result}\label{S:main}
It was noted in Sec. \ref{S:Introduction} that a 
Morris-Thorne wormhole can only be kept open by 
violating the null energy condition (NEC).  This 
condition states that for the energy-momentum tensor 
$T_{\alpha\beta}$, 
$T_{\alpha\beta}k^{\alpha}k^{\beta}\ge 0$ for all 
null vectors $T_{\alpha\beta}$.  In this section we 
are going to show that thanks to he extra spatial 
dimension, the wormhole throat can be lined with 
ordinary matter, while the violation of the NEC 
can be attributed to the existence of the fifth 
dimension.

To that end, we start with the four-dimensional 
null vector $(1,1,0,0)$, leaving the other null vectors 
for later.  The Einstein field equations in the orthonormal 
frame are 
\begin{equation}
   G_{\hat{\alpha}\hat{\beta}}=R_{\hat{\alpha}\hat{\beta}}-\frac{1}
{2}Rg_{\hat{\alpha}\hat{\beta}}=8\pi T_{\hat{\alpha}\hat{\beta}},
\end{equation}
where
\begin{equation}\label{E:metrictensor}
   g_{\hat{\alpha}\hat{\beta}}=
   \left(
   \begin{matrix}
   -1&0&0&0\\
   \phantom{-}0&1&0&0\\
   \phantom{-}0&0&1&0\\
   \phantom{-}0&0&0&1
   \end{matrix}
   \right).
\end{equation}
To simplify the notation, we now omit the hats.  So
$T_{00}=\rho$ is the energy density and $T_{11}=p_r$ 
is the radial pressure.  (The  NEC is therefore violated 
whenever $\rho+p_r<0.)$  We now have  
\begin{equation}\label{E:exotic}
   G_{00}+G_{11}=8\pi(\rho+p_r)=\left[R_{00}-\frac{1}{2}R(-1)
   \right]+\left[R_{11}-\frac{1}{2}R(1)\right]
   =R_{00}+R_{11}.
\end{equation}
Since we are primarily interested in the vicinity of the 
throat, we assume that $1-b(r_0)/r_0=0$.  We can now go 
directly to Appendix B and deduce that
\begin{equation}\label{E:aa}
  8\pi(\rho +p_r)|_{r=r_0}=R_{00}+R_{11}|_{r=r_0}=
  \frac{b'(r_0)-1}{r_0^2}
  \left[1+\frac{r_0}{2}\frac{\partial\mu(r_0,l)}{\partial r}
  \frac{1}{\mu(r_0,l)}\right].
\end{equation}
Recalling that $b'(r_0)<1$, it follows that for the 
four-dimensional case, $8\pi(\rho +p_r)|_{r=r_0}>0$  
whenever
\begin{equation}\label{E:bb}
   \frac{\partial\mu(r_0,l)}{\partial r}<0
\end{equation}
since $\mu(r_0,l)$ is assumed to be extremely small. 

For the five-dimensional case, consider the null 
vector $(1,0,0,0,1)$.  Then 
\begin{equation}
   8\pi T_{\alpha\beta}k^{\alpha}k^{\beta}=G_{00}
   +G_{44}=\left[R_{00}-\frac{1}{2}R(-1)\right]+
   \left[R_{44}-\frac{1}{2}R(1)\right]
   =R_{00}+R_{44}.
\end{equation}
Given that $1-b(r_0)/r_0=0$ again, we get
 \begin{equation}\label{E:cc}
    R_{00}+R_{44}|_{r=r_0}=
    \frac{1}{2}\frac{b'(r_0)-1}{r_0}\left[
    -\frac{\partial\Phi(r_0,l)}{\partial r}
    +\frac{\partial\mu(r_0,l)}{\partial r}
     \frac{1}{\mu(r_0,l)}\right].
\end{equation}
With inequality (\ref{E:bb}) in mind, 
suppose we also have 
\begin{equation}
   \frac{\partial\Phi(r_0,l)}{\partial r}<0.
\end{equation} 
Then $R_{00}+R_{44}|_{r=r_0}<0$ whenever
\begin{equation}
   \left |\frac{\partial\Phi(r_0,l)}{\partial r} \right|>
   \left |\frac{\partial\mu(r_0,l)}{\partial r}
   \frac{1}{\mu(r_0,l)}\right|.
\end{equation}
More precisely, from Eq. (\ref{E:aa}), if 
\begin{equation}
   \frac{\partial\mu(r_0,l)}{\partial r}\frac{1}{\mu(r_0,l)}
   <-\frac{2}{r_0}
\end{equation}
[which is consistent with (\ref{E:bb})], 
then $R_{00}+R_{44}|_{r=r_0}<0$ provided that  
\begin{equation}
   \frac{\partial\Phi(r_0,l)}{\partial r}=-A<
   \frac{\partial\mu(r_0,l)}{\partial r}\frac{1}{\mu(r_0,l)}.
\end{equation}
So the NEC is indeed violated at the throat in the 
five-dimensional case, even though it is met in the 
four-dimensional case.  

Summarizing, we have shown that the quantity
\begin{equation}
   \left |\frac{\partial\mu(r_0,l)}{\partial r}
   \frac{1}{\mu(r_0,l)}\right|
\end{equation}
is large enough to satisfy condition (\ref{E:aa}) 
(four-dimensional case) and  small enough to satisfy 
condition (\ref{E:cc}) (five-dimensional case). 

\section{The remaining conditions}
It was noted in Sec. \ref{S:main} that for the 
four-dimensional case, the NEC must also be satisfied for 
the null vectors $(1,0,1,0)$ and $(1,0,0,1)$.  To that end, 
we require that 
\begin{multline} 
   R_{00}+R_{22}|_{r=r_0}=R_{00}+ R_{33}|_{r=r_0}=
   -\frac{1}{2}\frac{\partial\Phi(r_0,l)}{\partial r}
   \frac{r_0b'(r_0)-b(r_0)}{r_0^2}+\frac{1}{2}
    \frac{r_0b'(r_0)-b(r_0)}{r_0^2}\\
     +\frac{b(r_0)}{r_0^3}=\frac{1}{2}\frac{b'(r_0)-1}{r_0}
     \left(1-\frac{\partial\Phi(r_0,l)}{\partial r}\right)
     +\frac{1}{r_0^2}>0.
\end{multline}
This condition is met provided that $b'(r_0)$ is 
sufficiently close to unity.

\section{Conclusion}

It is well known that a Morris-Thorne wormhole can only be 
sustained by violating the null energy condition, requiring 
the existence of ``exotic matter."  It has been shown that 
this requirement can be met by assuming the existence of 
an extra spatial dimension: the throat of the wormhole can 
be lined with ordinary matter, while the extra spatial 
dimension is responsible for the unavoidable energy 
violation.

The proposed line element is
\begin{equation*}
  ds^2=-e^{2\Phi(r,l)}dt^2 +e^{2\lambda(r,l)}dr^2+r^2
  (d\theta^2+\text{sin}^2\theta\,d\phi^2)+[\mu(r,l)]^2 dl^2,
\end{equation*}
where $l$ is the extra coordinate.  It is subsequently shown 
that to be consistent with string theory, we must have 
$\partial\Phi(r,l)/\partial l\equiv0$, while the shape function 
must have the usual form for a Morris-Thorne wormhole.

The requirements regarding the energy conditions can be 
summarized as follows: in the four-dimensional case [null 
vector $(1,1,0,0)$],
\begin{equation}\label{E:condition1}
  8\pi(\rho +p_r)|_{r=r_0}=R_{00}+R_{11}|_{r=r_0}=
  \frac{b'(r_0)-1}{r_0^2}
  \left[1+\frac{r_0}{2}\frac{\partial\mu(r_0,l)}{\partial r}
  \frac{1}{\mu(r_0,l)}\right]>0
\end{equation}
whenever
\begin{equation}\label{E:condition1a}
   \frac{\partial\mu(r_0,l)}{\partial r}<0
\end{equation}
since $\mu(r_0,l)$ is assumed to be extremely small.

In the five-dimensional case [null vector $(1,0,0,0,1)]$,
suppose we have 
\begin{equation}\label{E:one} 
  \frac{\partial\Phi(r_0,l)}{\partial r}<0.  
\end{equation}
Then   
 \begin{equation}\label{E:condition2}
    R_{00}+R_{44}|_{r=r_0}=
    \frac{1}{2}\frac{b'(r_0)-1}{r_0}\left[
    -\frac{\partial\Phi(r_0,l)}{\partial r}
    +\frac{\partial\mu(r_0,l)}{\partial r}
     \frac{1}{\mu(r_0,l)}\right]<0
\end{equation}
whenever
\begin{equation}\label{E:two}
   \frac{\partial\mu(r_0,l)}{\partial r}
   \frac{1}{\mu(r_0,l)}<-\frac{2}{r_0}
\end{equation}
and
\begin{equation}\label{E:three}
   \frac{\partial\Phi(r_0,l)}{\partial r}=-A<
   \frac{\partial\mu(r_0,l)}{\partial r}\frac{1}{\mu(r_0,l)}.
\end{equation}
In summary, conditions (\ref{E:one}), (\ref{E:two}), and 
(\ref{E:three}) imply that the NEC is met in the 
four-dimensional case and that the unavoidable violation of 
the NEC can be attributed to the fifth dimension.

To finish our discussion, we need to recall that for the  
shape function $b=b(r)$ at the throat $r=r_0$, $b'(r_0)$ 
must be sufficiently close to unity.

It should also be emphasized that the requirement that 
$\mu(r,l)$ has a small magnitude is consistent with string 
theory.

\textbf{APPENDIX A\quad The curvature 2-forms}

\begin{multline} \Omega^0_{\phantom{1}1}=\left [-\frac{\partial^2\Phi(r,l)}
   {\partial r^2}\left (1-\frac{b(r)}{r}\right )
   -\left (\frac{\partial\Phi(r,l)}{\partial r}\right )^2
   \left (1-\frac{b(r)}{r}\right )\right.\\ \left.
   +\frac{1}{2}\frac{\partial\Phi(r,l)}{\partial r}
   \frac{rb'(r)-b(r)}{r^2}\right ]\theta^0\wedge\theta^1
   \end{multline}
\begin{equation}
   \Omega^0_{\phantom{1}2}=-\frac{1}{r}\frac{\partial\Phi(r,l)}
   {\partial r}\left (1-\frac{b(r)}{r}\right)\theta^0\wedge\theta^2   
\end{equation}
\begin{equation}
   \Omega^0_{\phantom{1}3}=-\frac{1}{r}\frac{\partial\Phi(r,l)}
   {\partial r}\left (1-\frac{b(r)}{r}\right)\theta^0\wedge\theta^3
\end{equation}
\begin{equation}
   \Omega^0_{\phantom{1}4}=-\frac{\partial\Phi(r,l)}{\partial r}
   \frac{\partial\mu(r,l)}{\partial r}\left(1-\frac{b(r)}{r}\right )
   \frac{1}{\mu(r,l)}\theta^0\wedge\theta^4
\end{equation}
\begin{equation}
   \Omega^1_{\phantom{1}2}=\frac{1}{2}\frac{rb'(r)-b(r)}{r^3}
   \theta^1\wedge\theta^2
\end{equation}   
\begin{equation}
   \Omega^1_{\phantom{1}3}=\frac{1}{2}\frac{rb'(r)-b(r)}{r^3}
   \theta^1\wedge\theta^3
\end{equation}   
\begin{equation}
   \Omega^1_{\phantom{1}4}=\left [-\frac{\partial^2\mu(r,l)}
   {\partial r^2}\left(1-\frac{b(r)}{r}\right)+\frac{1}{2}
   \frac{\partial\mu(r,l)}{\partial r}\frac{rb'(r)-b(r)}{r^2}
   \right ]\frac{1}{\mu(r,l)}\theta^1\wedge\theta^4    
\end{equation} 
\begin{equation}
   \Omega^2_{\phantom{1}3}=\frac{b(r)}{r^3}
   \theta^2\wedge\theta^3 
\end{equation}
\begin{equation}
   \Omega^2_{\phantom{1}4}=-\frac{1}{r}\frac{\partial\mu(r,l)}{\partial r}
   \left (1-\frac{b(r)}{r}\right )\frac{1}{\mu(r,l)}
   \theta^2\wedge\theta^4
\end{equation}
\begin{equation}
   \Omega^3_{\phantom{1}4}=-\frac{1}{r}\frac{\partial\mu(r,l)}{\partial r}
   \left(1-\frac{b(r)}{r}\right )\frac{1}{\mu(r,l)}\theta^3\wedge\theta^4 
\end{equation}
\textbf{APPENDIX B\quad The components of the Ricci tensor}

\begin{multline}\label{E:R00}
 R_{00}= -\frac{1}{2}\frac{\partial\Phi(r,l)}{\partial r}\frac{rb'(r)-b(r)}{r^2}
 +\frac{\partial^2\Phi(r,l)}{\partial r^2}\left(1-\frac{b(r)}{r}\right)
+\left(\frac{\partial\Phi(r,l)}{\partial r}\right)^2
 \left(1-\frac{b(r)}{r}\right)
\\+\frac{2}{r}\frac{\partial\Phi(r,l)}
 {\partial r}\left(1-\frac{b(r)}{r}\right) 
 +\frac{\partial\Phi(r,l)}{\partial r}\frac{\partial\mu(r,l)}{\partial r}
 \left(1-\frac{b(r)}{r}\right)\frac{1}{\mu(r,l)},
\end{multline}
\begin{multline}\label{E:R11}
 R_{11}=\frac{1}{2}\frac{\partial\Phi(r,l)}{\partial r}
 \frac{rb'(r)-b(r)}{r^2}-\frac{\partial^2\Phi(r,l)}{\partial r^2}\left
 (1-\frac{b(r)}{r}\right)-\left(\frac{\partial\Phi(r,l)}{\partial r}\right)^2
 \left(1-\frac{b(r)}{r}\right)\\+\frac{rb'(r)-b(r)}{r^3}-\left[
 \frac{\partial^2\mu(r,l)}{\partial r^2}\left(1-\frac{b(r)}{r}\right)
 -\frac{1}{2}\frac{\partial\mu(r,l)}{\partial r}\frac{rb'(r)-b(r)}{r^2}\right]
 \frac{1}{\mu(r,l)}
\end{multline}

\begin{multline}
 R_{22}=R_{33}=-\frac{1}{r}\frac{\partial\Phi(r,l)}{\partial r}\left(1-\frac{b(r)}{r}
 \right)+\frac{1}{2}\frac{rb'(r)-b(r)}{r^2}+\frac{b(r)}{r^3}\\-\frac{1}{r}\left(1-\frac{b(r)}{r}\right)
 \frac{\partial\mu(r,l)}{\partial r}\frac{1}{\mu(r,l)},
\end{multline}
\begin{multline}
 R_{44}=-\frac{\partial\Phi(r,l)}{\partial r}\frac{\partial\mu(r,l)}{\partial r}
 \left(1-\frac{b(r)}{r}\right)\frac{1}{\mu(r,l)}\\-\left[\frac{\partial^2\mu(r,l)}{\partial r^2}
 \left(1-\frac{b(r)}{r}\right)-\frac{1}{2}\frac{\partial\mu(r,l)}{\partial r}\frac{rb'(r)-b(r)}{r^2}
 \right]\frac{1}{\mu(r,l)}\\-\frac{2}{r}\left(1-\frac{b(r)}{r}\right)\frac{\partial\mu(r,l)}
 {\partial r}\frac{1}{\mu(r,l)}.
\end{multline}

\end{document}